# Teaching Sustainable Creative Technologies

Three methods for more carbon aware creative production.

Chelsea Thompto
School of Visual Art
Virginia Tech
Blacksburg, VA, USA
cthompto@vt.edu

## ABSTRACT

Artists and especially new media artists contribute to public perceptions and adoption of new technologies through their own use of emerging media technologies such as augmented and virtual reality, generative image systems, and high-resolution displays in the production of their work. In this way, art and media production can be understood as part of the larger issue of unsustainable computational consumption. As such, it is critical for artists to develop, share, and promote new and more sustainable methods of engaging with technology, especially within the context of higher education. This paper will explore how artists might implement more sustainable methods by considering the relationship between the technical approaches of compute reuse, sustainable web development, and frugal computing, and the concepts of material specificity[1], futurity, and media archaeology[2]. Proposing three methods of less carbon-intensive artistic production and a set of guidelines for introducing sustainable methods into arts and technology curriculum, this paper will outline not only the technical viability of these approaches but also the rich conceptual opportunities these approaches might offer to artists and viewers alike. For each method, models for pedagogical implementation will be explored with an emphasis on how local resources and sustainability contexts should play a role.

## INTRODUCTION

Digital artistic production is closely linked to carbon emissions, and at the same time art is a fundamental part of human experience. We can no more stop making art than we can stop making music, or poetry, or scientific discoveries. Yet as a society, we must all be working to reduce our carbon emissions. So, it is critical that as we train the future generations of creative technologists and digital artists, we do so while centering emissions and our changing climate.

There is an abundance of examples of low carbon methods within digital art and new media, from solar internet projects like *Solar Protocol* [1], to theories like Permacomputing Aesthetics [2], to movements like Web Revival [3], events like Small File Media Festival [4] and more. However, the problem is not that digital artists have yet to engage with computing within limits, but rather that these approaches remain on the very edges of creative technological practice and have not been widely adopted within digital creative production pedagogies at the university level where curriculum often favors "industry standard" and carbon intensive production pipelines[3].

Artists, and perhaps especially new media artists, contribute to public perception and adoption of new technologies through their use of emerging media technologies such as augmented and virtual reality, generative image systems, and high-resolution displays in the production of their work. In this way, art and media production are mechanisms within the larger machine of unsustainable computational consumption. As such, it is critical for artists to develop, share, and promote new and more sustainable methods of engaging with technology, especially within the context of higher education. This context can be narrowed to two central questions:

1. How can we teach the technical and conceptual skills needed for (new media) artists to acknowledge our climate context in their work?

2. How can digital arts education play a role in resisting the aestheticization and adoption of carbon intensive[4] digital technologies?

---

[1] In art, material specificity refers to the idea that all materials have inherited and culturally situated meanings that artists can use and manipulate in the production of their work.
[2] Media archaeology is the study of media history through lesser-known devices and processes.

[3] While a systematic review of programs has not been undertaken, current new media and digital art references, textbooks, and other curriculum materials rarely cover this subject.
[4] Including but not limited to: ever increasing screen resolutions, faster frame and refresh rate displays, "AI" power graphics technologies, always online digital experiences, and more.

Answering these questions brings us quickly to three main challenges that exist in addressing these issues within the classroom:

1. Climate change touches all aspects of artistic labor, from ideation to production to dissemination. This ubiquity makes it difficult to address in a classroom setting or really in one's practice at all.

2. While examples exist, as stated above and as will be touched on within each method, many approaches to low carbon creative computing are technologically complex. This complexity can make introducing these topics to beginners very challenging, especially when the context is not computer science but digital art/creative technologies.

3. There is a lack of artworks which engage low carbon methods while not being about low carbon methods. In other words, there is a need for showing how to do creative work sustainably when the work itself isn't necessarily about sustainability.

To address these challenges, educators must identify actionable areas of improvement within the technologies they are teaching and craft projects and hands-on experiences and weave low carbon thinking into the aesthetic, theoretical, and technical content of the course.

While the scope of artistic work does not usually come at a scale that would see large gains in overall emissions reductions, it is nonetheless important for artists to engage with technologies with the environment in mind. Doing so allows for:

- Technological limits to beneficially inform aesthetic choices.

- Artists to challenge the preconceived notions of technological standards and norms.

- The technological substrate of the work to be understood as highly material rather than as "in the cloud" or immaterial.

- Artists to help enfold sustainable practices into their imaginaries.

Each of the methods that follow link technical and conceptual methods with an existing artwork and outline a range of classroom integrations of varying technical complexity.

## METHOD 1 – COMPUTE REUSE AS MEDIA ARCHAEOLOGY

An exploration of how reusing obsolete or out-moded hardware (e-waste) can simultaneously extend the life of compute equipment while also enabling artists to include the social context for which the equipment was initially made in the conceptual development of their work. This approach can be seen in the work of PAMAL-Art and their work *Profound Telematic Time (P.T.T.)* [5], 2022, which makes use of and explores the culture and history surrounding the Minitel[5] network.

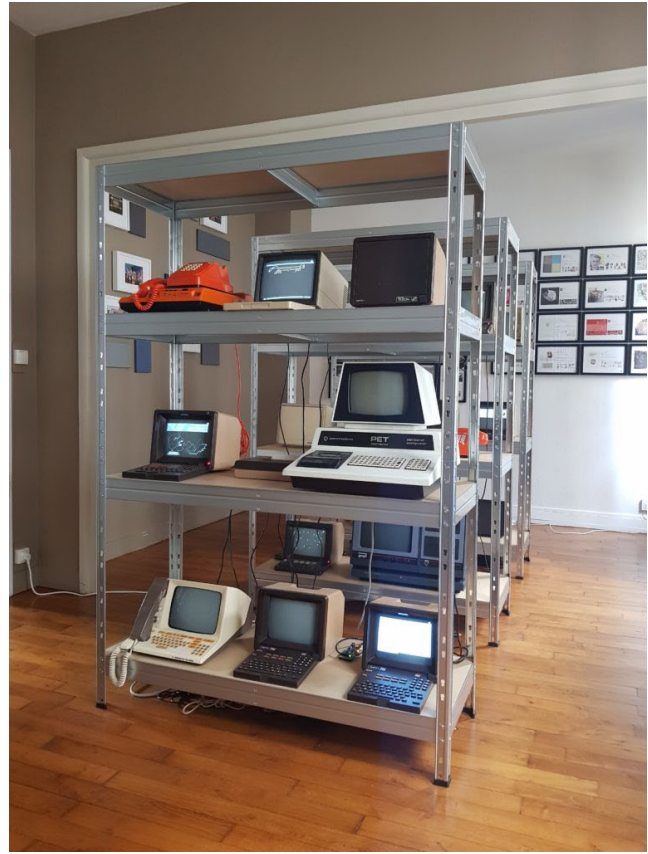

**Figure 1: Installation view of *Profound Telematic Time (P.T.T.)***

Classroom engagement with this method might include a site visit to a local e-waste facility and a project wherein students explore the history of and develop a new work for a specific piece of "old hardware". In a coding intensive or sufficiently advanced undergraduate course, this could either be done via emulation of the hardware or acquisition of the original hardware depending on class scope and prior experience. In a course that does not focus on coding or where students lack the technical capacity for such work, they could instead adopt the limits of their chosen hardware (color palette, resolution, etc.) in the development of a new artwork. One of the most basic examples of this is to have students take photos and then convert them using the default dithering and compression algorithms of early home computers such as the Commodore 64[6].

---

[5] The Minitel network was an online service in the 1980s which predated the modern internet.

The goal of this method is to foster a mindset of reuse as both ecologically conscientious and conceptually rich, while also introducing students to the process of understanding and developing new works within the constraints of older technology.

## METHOD 2 – SUSTAINABLE WEB DEVELOPMENT [7] AND MATERIAL SPECIFICITY

An exploration of how designing websites with a low carbon mindset on the backend can be brought to frontend development in the form of low resolution/dithered images, limited color palettes/animations, etc. to highlight these choices to end users/viewers. The work *Solar Protocol* [1] by Tega Brain, Alex Nathanson, and Benedetta Piantella exemplifies this holistic approach to sustainable backend and frontend creative production. The *Low-Tech Magazine* [6] website is another excellent example of this ethos[8]. These projects can help introduce students to the (often invisibilized) materiality of our networked experiences.

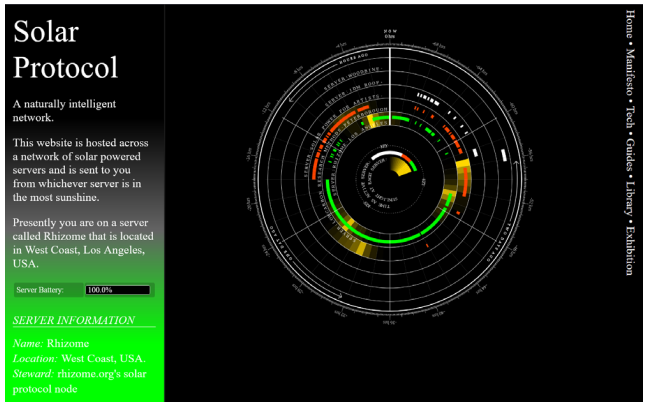

**Figure 2: A screenshot of the *Solar Protocol* website featuring server battery and location information.**

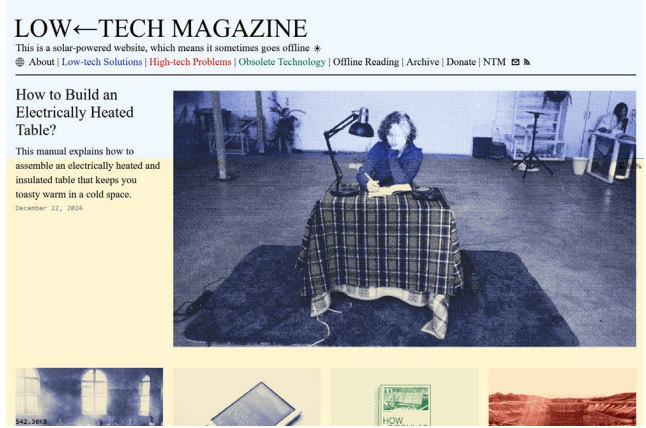

**Figure 3: A screenshot of the *Low-Tech Magazine* website featuring server battery and dithered images.**

In the classroom, this could take the form of facilitating hands-on experience with network hardware (creating simple servers using microcomputers or repurposed machines) and challenging students to work within visual and technical limits in the production (coding and deployment) of a website. In the absence of hands-on hardware to engage with, or in the context where such explorations are too complex, students can be guided to website carbon emissions calculators such as Website Carbon [7] to help them understand the impact of their work and to critically interrogate existing web design.

For students lacking coding experience, HTML and CSS provide both an excellent introduction to code and access to web publications that can and should be treated as a destination for creative expression rather than mere on-ramp to the more carbon intensive world of JavaScript. In this way, sustainable web development can be used to frame introductions to web development more broadly. I have been using this framing in my web development lessons, within the context of creative coding curriculum, since 2021[9].

The goal of this method is to help students understand the materiality of the internet and to reframe HTML and CSS (as well as other lightweight web processes) as complete approaches to web development rather than mere steppingstones to more carbon intensive processes.

---

[6] In my classes, I use https://doodad.dev/dither-me-this/ as quick and easy tool to get students exploring this concept.
[7] Web development here refers to both frontend and backend development and is reflective of the need for artists and designers to engage with both sides in the development of truly ecologically conscious work.
[8] Of particular note with *Low Tech Magazine* is that fact that the original solar website, designed in and implemented in 2018, was redesigned and redeployed in 2023. This revision process provides critical insights to the work of sustaining a solar powered server over a long period of time. A summary of that work can be found here: https://solar.lowtechmagazine.com/2023/06/rebuilding-a-solar-powered-website/
[9] I wrote the following assignment template for my students to get them introduced to sustainability web design with HTML and CSS: https://github.com/cthompto/Sustainable-and-Accessible-Web-Design-with-HTML-and-CSS

## METHOD 3 – FRUGAL COMPUTING IN ARTIST CONCEPTIONS OF THE FUTURE

An exploration of how artists can develop work in ways that challenge the "always on" approach in contemporary gallery/museum display settings and the misconception of digital as immaterial to instead imagine projects whose production and display are tied to available resources. The work of *Energy Transition Design* [8] to produce solar power systems for art and design contexts offers ways to imagine artworks that respond to the availability of renewable energy.

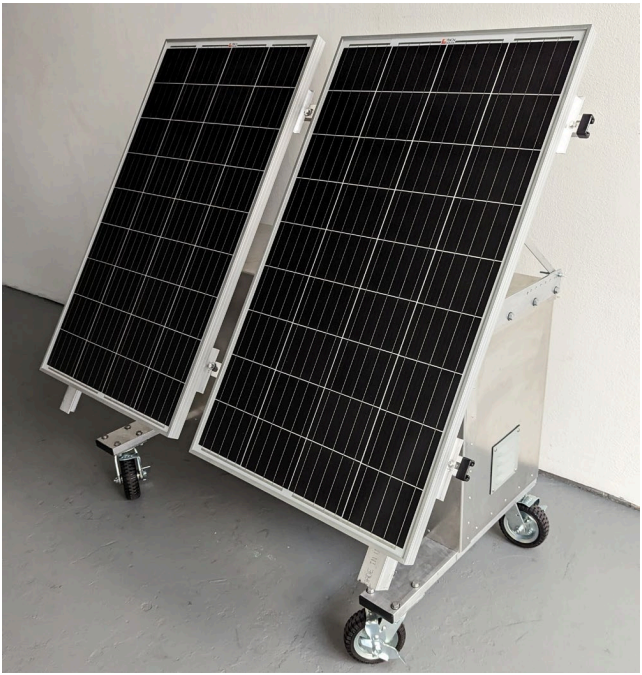

**Figure 4: A solar cart designed by Energy Transition Design.**

The classroom implementation of this could involve students redesigning an existing exhibition space to be responsive to energy availability and then designing future projects for the redesigned space. In the case of my recent climate class[10], this method was enacted through the co-development of a renewables powered autonomous display system, inspired by simple solar power displays such as those used for road signs, over the course of the semester. Students had the opportunity to see how the process develops from a proposal to the development of a working prototype and had the opportunity to imagine work for the system as part of the class's culminating experience[11].

The goal of this method is to foster sustainable thinking into imaginings of the future and to help students gain the tools and insights needed to challenge conventional wisdom and methodologies around the distribution and display of digital artwork.

## STUDENT PRECEPTIONS

While the work of building student and institutional capacity to more fully engage with the first outlined method is still in progress, I have had the opportunity to introduce dithering and other methods of image/data compression to students. Though a small sample size (4 students), all students reported finding the process and its results to be aesthetically appealing and technically interesting. The following example shows a student photograph and one of four dithered versions they created. This student stated they were particularly drawn to the color palette and the horizontal line distortions created by the dithering algorithm.

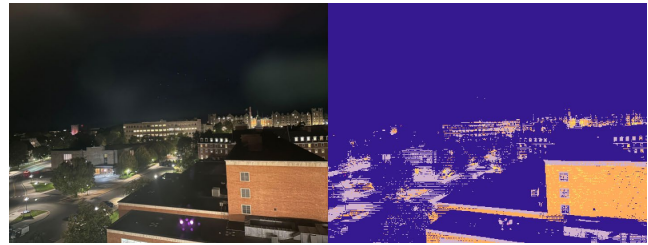

**Figure 5: A student photograph (left) and their dithered version (right).**

Response to method 2 (sustainable web development), which I have taught in a variety of formats over the past 4 years, has been generally positive. Three key pieces of reflection/feedback from these units (compiled using electronic feedback surveys) have been:

- Surprise at how resource intensive even a seemingly simple website is on the contemporary web.
- A higher level of interest in HTML and CSS as languages/tools in their own right (and not merely as a steppingstone to more "advanced" tools or languages) compared to prior web development projects which did not emphasize sustainability.
- Those with a self-reported interest in sustainability appreciated being able to connect that interest to their university coursework.

Throughout my time teaching this content, I have found that students with a wide range of visual styles and professional goals have been able to connect positively to these methods of working. The following example shows a student who wanted to create a minimalist design portfolio. The project challenged students to use only HTML and CSS and to use efficient media formats. The resulting home page manages to come across as professional while maintaining an "A+" rating on the Website Carbon Calculator[12].

---

[10] In Spring 2025 I am teaching a combination graduate and undergraduate course titled "Creative Technologies During Climate Crisis" at Virginia Tech.

[11] This research is being funded through a Faculty Initiated Research Grant (FIRG) from Virginia Tech's College of Architecture, Arts, and. Design.

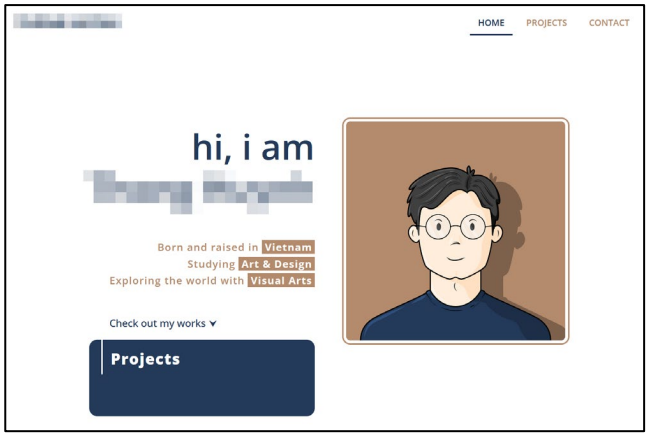

Figure 6: A screenshot of a student's sustainably designed website (name blurred in top left and center for student privacy).

Overall, the student response to introducing sustainable methods in creative technology contexts has been very positive. Nearly all students report that they are concerned about climate change and have a desire to learn more about how it might impact their lives and work in the future. While some students struggled with the technical aspects of these assignments, in my own classroom experiences, it has not been at a higher rate compared to other, more conventional, technical approaches to these subjects.

## IMLPEMENTATION GUIDELINES

Returning to the first two questions:

- How can we teach the technical and conceptual skills needed for (new media) artists to acknowledge our climate context in their work?
- How can digital arts education play a role in resisting the aestheticization of digital maximalism?

And considering the context, examples, and students' experiences outlined, I propose the following guidelines for addressing these issues in the classroom. These guidelines seek to provide a set of considerations that can be applied beyond the three methods given earlier in this paper to any initiative aimed at making art and technology education more sustainable.

1. Be mindful of your location. In this case, meaning both the specific site in which the project is taking place (the lesson or project) and the sites that it may be exhibited or shared. Taking this context into consideration allows for:

    - The identification of potential local partners (e-waste facilities, renewable energy sites, community groups, etc.).
    - An understanding of how categories such as "renewable" or "sustainable" might mean different things in different contexts.
    - The exhibition of the work to be considered in its overall ecological impact and life span.

2. Give students multiple ways into the subject and opportunities to give feedback. Students will come to this work with different levels of familiarity about both the broader subject of climate/sustainability and the specific arts-based technical and conceptual knowledge being covered. Allow students to draw on their prior knowledge and lived experience when facing new material. Taking this into consideration empowers students to:

    - Draw connections between coursework and their lived experiences.
    - Shape coursework by allowing them to share their concerns and interests.
    - Move through the paralyzing feeling that climate change is too big of a subject to face by focusing on what they do know and what they can do/learn.

3. Consider the learning outcomes of replaced technologies and skills in a class. If a certain technology is being replaced in a course for something more sustainable (perhaps coding with a retro game engine instead of unity) it is important to consider how the more sustainable option can address similar core concepts. Doing so will:

    - Demonstrate (to both students and colleagues) that working sustainability does not mean sacrificing rigor or core learning objectives.
    - Allow the course and project to continue to fulfill its role in the larger context of the workshop, class, or degree.

4. Provide real world examples and hands-on experiences. Sustainable practices often fall prey to the idea that they are new or untested and thereby less suited for the classroom. Providing students with tangible examples can:

    - Demonstrate that sustainable practices are already in use and are often a viable alternative.

---

[12] I used the calculator at https://www.websitecarbon.com/ to rate the student's work.

- Give students a more meaningful and holistic understanding of how these methods look and feel in practice.
5. And finally, iterate. Art, technology, and sustainability are all shifting and evolving fields. One must be willing to change and continually develop their coursework to meet the moment.

## CONCLUSION

Taken together, these methods and guidelines begin to outline new modes of teaching artistic production that are responsive to our shared climate crisis as opposed to complacent in driving digital consumption. It is critical that such methods be enfolded into new media art education across a broad spectrum of technical processes including but not limited to installation art, video and animation production, web development, and interactive media/game development. Further, these methods not only offer low carbon alternatives to traditional media art production, but they also offer new ways of thinking about technology which will be of critical importance for future generations of creative practitioners.

These methods offer new ways for those of us teaching creative technologies to do so in more sustainable ways and begin to integrate low carbon methods into all aspects of our curriculum. By introducing students to the work of existing practitioners and by drawing connections between technical approaches and conceptual concerns, we can foster a perception of these practices that sees them as an opportunity rather than a burden to the creative process. Again, while the scope of individual artistic works is normally rather limited in a technical sense, artists drive interest in and adoption of technologies at a social and cultural level and must therefore expand their work to help imagine and enact new ways of thinking and making sustainably within the arts and beyond.

## KEYWORDS

Materiality, new media art, design, media archaeology, e-waste, sustainable web development, frugal computing, carbon awareness.